\documentclass[%
 reprint,
superscriptaddress,
 amsmath,amssymb,
 aps,
prb,
floatfix,
]{revtex4-1}

\usepackage{graphicx}% Include figure files
\usepackage{dcolumn}% Align table columns on decimal point
\usepackage{bm}% bold math
\usepackage[colorlinks=true,allcolors=blue]{hyperref}% add hypertext capabilities
\usepackage{array,booktabs,tabularx}
\usepackage[caption=false]{subfig}
\usepackage[USenglish]{babel}
\usepackage{braket}
\usepackage{xcolor}
\usepackage[normalem]{ulem}
\usepackage{notes2bib}

\begin{document}

\title{Phonon collapse and van der Waals melting of the 3D charge density wave of VSe$_2$}
%\title{Phonon softening and van der Waals melting of the 3D charge density wave of VSe$_2$}

\author{Josu Diego}
\affiliation{Centro de Física de Materiales (CSIC-UPV/EHU), San Sebastián, Spain}

\author{A. H. Said}
\affiliation{Advanced Photon Source, Argonne National Laboratory, Lemont, IL 60439}

\author{S. K. Mahatha}
\affiliation{Ruprecht Haensel Laboratory, Deutsches Elektronen-Synchrotron DESY, 22607 Hamburg, Germany}

\author{Raffaello Bianco}
\affiliation{Centro de Física de Materiales (CSIC-UPV/EHU), San Sebastián, Spain}
             
\author{Lorenzo Monacelli}
\affiliation{Dipartimento di Fisica, Universit\`a di Roma La Sapienza,  Roma, Italy}
\affiliation{Graphene Labs, Fondazione Instituto Italiano di Tecnologia, Italy}

\author{Matteo Calandra}
\affiliation{Dipartimento di Fisica, Università di Trento, Via Sommarive 14, 38123 Povo, Italy.}
\affiliation{Sorbonne Universit\'es, CNRS, Institut des Nanosciences de Paris, UMR7588, F-75252, Paris, France}
\affiliation{Graphene Labs, Fondazione Instituto Italiano di Tecnologia, Italy}

\author{Francesco Mauri}
\affiliation{Dipartimento di Fisica, Universit\`a di Roma La Sapienza, Roma, Italy} 
\affiliation{Graphene Labs, Fondazione Instituto Italiano di Tecnologia, Italy}

\author{K. Rossnagel}
\affiliation{Ruprecht Haensel Laboratory, Deutsches Elektronen-Synchrotron DESY, 22607 Hamburg, Germany}
\affiliation{Institut f\"{u}r Experimentelle und Angewandte Physik, Christian-Albrechts-Universit\"{a}t zu Kiel, 24098 Kiel, Germany}

\author{Ion Errea}
\email{ion.errea@ehu.eus}
\affiliation{Centro de Física de Materiales (CSIC-UPV/EHU), San Sebastián, Spain}
\affiliation{Fisika Aplikatua 1 Saila, Gipuzkoako Ingeniaritza Eskola, University of the Basque Country (UPV/EHU), San Sebastián, Spain}
\affiliation{Donostia International Physics Center (DIPC), San Sebastián, Spain}

\author{S. Blanco-Canosa}
\email{sblanco@dipc.org}
\affiliation{Donostia International Physics Center (DIPC), San Sebastián, Spain}
\affiliation{IKERBASQUE, Basque Foundation for Science, 48013 Bilbao, Spain}

\date{May 2020}
\begin{abstract}
Among transition metal dichalcogenides (TMDs), VSe$_2$ is considered to develop a purely 3-dimensional (3D) charge-density wave (CDW) at T$_{CDW}$=110 K. Here, by means of high resolution inelastic x-ray scattering (IXS), we show that the CDW transition is driven by the collapse of an acoustic mode at the critical wavevector \textit{q}$_{CDW}$= (2.25 0 0.7) r.l.u. and critical temperature T$_{CDW}$=110 K. The softening of this mode starts to be pronounced for temperatures below 2$\times$ T$_{CDW}$ and expands over a rather wide region of the Brillouin zone, suggesting a large contribution of the electron-phonon interaction to the CDW formation. This interpretation is supported by our first principles calculations that determine a large momentum-dependence of the electron-phonon interaction, peaking at the CDW wavevector, in the presence of nesting. Fully anharmonic {\it ab initio} calculations confirm the softening of one acoustic branch at \textit{q}$_{CDW}$ as responsible for the CDW formation and show that van der Waals interactions are crucial to melt the CDW. Our work also highlights the important role of out-of-plane interactions to describe 3D CDWs in TMDs.  
\end{abstract}

%\begin{document}
\maketitle

%\section{Introduction}
A comprehensive and detailed understanding of electronic ordering and charge-density-wave (CDW) formation is attracting great efforts in condensed matter physics \cite{Gru94}. In particular, its dynamical nature is the focus of strong debate in correlated oxides and high T$_c$ superconducting cuprates \cite{Fra20}, where fluctuations of the charge order parameter \cite{Ar19}, dispersive CDW excitations \cite{Chaix17}, and phonon anomalies are observed \cite{LeTac14}. Microscopically, the subtle balance between electron-phonon interaction (EPI) and  nested portions of the Fermi surface (singularities in the electronic dielectric function, $\chi_q$, at \textit{q}$_{CDW}$= 2\textit{k}$_F$) determines the origin and stabilization of the charge periodicities \cite{Chan73}. While the Fermi surface nesting scenario survives for 1D and quasi-1D systems (Peierls transition), its role in higher dimensions remains largely questioned \cite{Miao19,PhysRevB.77.165135}. 

Among the solids showing electronic charge ordering, layered transition metal dichalcogenides (TMDs) represent the first crystalline structures where 3D CDWs were discovered \cite{Wil74}. 1\textit{T}-VSe$_2$ (space group \textit{P$\overline{3}$m1}) belongs to the series of layered TMDs that develops a 3D-CDW as a function of temperature, T$_{CDW}$= 110 K. However, unlike the isostructural 1\textit{T}-TiSe$_2$, which adopts a commensurate 2$\times$2$\times$2 CDW ordering with \textit{q}$_{CDW}$=(0.5 0 0.5) \cite{Joe2014}, 1\textit{T}-VSe$_2$ develops a more complex incommensurate 3D pattern in its CDW phase with a \textit{q}$_{CDW}$=(0.25 0 -0.3) CDW wavevector \cite{Ross11}. 1\textit{T}-VSe$_2$ is rather unique among the 1\textit{T}-polytypes because it develops anomalies in its transport properties and magnetic susceptibility \cite{Brug76} that more closely resemble those of 2\textit{H}-polytypes: e.g. T$_{CDW}$[2\textit{H}-NbSe$_2$]= 33 K, T$_{CDW}$[2\textit{H}-TaSe$_2$]= 122 K and presents the lowest onset temperature among them, i.e., T$_{CDW}$[1\textit{T}-TiSe$_2$]= 200 K, T$_{CDW}$[1\textit{T}-TaS$_2$]= 550 K \cite{Ross11}. The sizable difference between T$_{CDW}$[1\textit{T}-VSe$_2$] and its 1\textit{T} counterparts can be attributed to the occurrence of large fluctuation effects that lower the mean-field transition temperature \cite{Lee73} or to the out-of-plane coupling \cite{Driza2012} between neighboring VSe$_2$ layers assisted by the weak short-range van der Waals interactions \cite{Lin2019}. Moreover, the theoretical input based on \textit{ab initio} calculations is also limited for all these TMDs undergoing CDW transitions due to the breakdown of the standard harmonic approximation for phonons, which cannot explain the stability of the high-temperature undistorted phases \cite{PhysRevB.92.140303}. This hinders the study of both the origin and the melting of the electronically modulated state, complicating the comprehensive understanding of the CDW formation. 

%In such a case, the ordering with finite correlation length persists well above the transition temperature, suggesting that the ground state of VSe$_2$ is governed by electron-electron correlation effects. In addition, the disagreement between experiments and theory in TMDs, and even in simple metals, has its origin in the basic harmonic approach, which only takes into account the first term in the Taylor expansion of the lattice potential. Therefore, the electron-phonon interaction that affects the electronic states distorting the band structure close to the Fermi surface is completely neglected.

\begin{figure*}
\begin{center}
\includegraphics[width=2.0\columnwidth,draft=false]{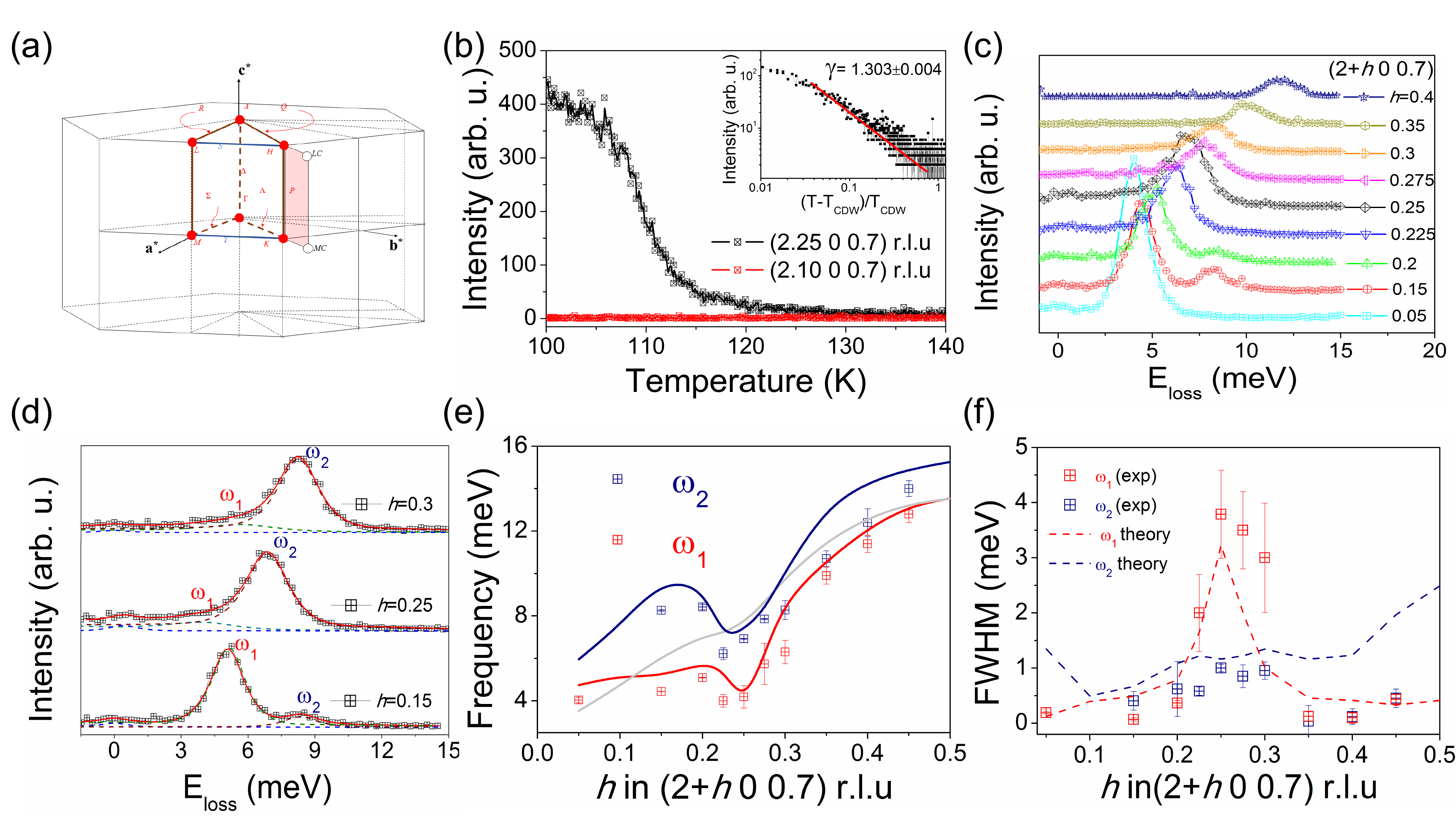}
\caption{(a) The \textit{k}-vector types of space group  \textit{P$\overline{3}$m1} (164) \cite{Aroyo:xo5018}.  (b) Temperature dependence of the elastic line at (2.25 0 0.7) r.l.u. showing the onset of the CDW at 110 K. Inset, scaling analysis of the elastic peak. (c) Energy-momentum dependence of the low energy acoustic phonons at 300 K from 0.05$<$\textit{h}$<$0.4 r.l.u. along the (2+\textit{h} 0 0.7) direction. The spectra are vertically off-set for clarity. (d) Close-up view of the phonon fitting at 300 K for selected momentum transfers, identifying the $\omega_1$ and $\omega_2$ phonons. (e) Experimental (points) and calculated (solid lines) dispersion of the low energy acoustic phonons at 300 K. The grey line stands for the silent mode, not observed experimentally. (f) Momentum dependence of the linewidth for $\omega_1$ and $\omega_2$ obtained from the fitting of the experimental spectra to damped harmonic oscillators. The calculated linewidth including the contribution of the electron-phonon interaction and anharmonicity is shown as dashed lines.}
\label{Fig1}
\end{center}
\end{figure*}

From the electronic point of view, angle resolved photoemission (ARPES) experiments in VSe$_2$ reported asymmetric dogbone electron pockets centred at \textit{M(L)} \cite{Stro12} that follow the threefold symmetry of the Brillouin zone (BZ) interior, with nesting vectors closely matching those observed by x-ray scattering \cite{Tsu82}. The formation of the CDW results from the 3D warping of the Fermi surface in the \textit{ML} plane (fig. 1a shows the high-symmetry points of the Brillouin zone of the hexagonal lattice of VSe$_2$). Moreover, photoemission data also find a partial suppression of the density of states near \textit{E}$_F$ on the nested portion below 180 K, indicating that a pseudogap opens at the Fermi surface \cite{Ter03}. However, a detailed investigation of the electronic structure is complicated by the 3D nature of the CDW order, and the momentum dependence of the EPI and the response of the lattice to the opening of the gap at E$_F$ remains unsolved. In fact, inelastic x-ray scattering (IXS) and theoretical calculations discarded the Fermi surface nesting scenario proposed for 2\textit{H}-NbSe$_2$ \cite{Web11Nb,PhysRevB.80.241108} and 1\textit{T}-TiSe$_2$ \cite{Web11Ti,PhysRevLett.106.196406} and emphasized the critical role of the momentum dependence of the electron-phonon interaction. In addition, it has been recently demonstrated that large anharmonic effects are required to suppress the CDW phases in TMDs and understand their phase diagrams \cite{PhysRevB.92.140303,doi:10.1021/acs.nanolett.9b00504,zhou2019anharmonic,bianco2020weak}. 

Here, we report the temperature dependence of the soft phonon mode in VSe$_2$ by high resolution IXS. We show that a low energy acoustic branch at \textit{q}$_{CDW}$ undergoes a softening of $\approx$3 meV from 2$\times$ T$_{CDW}$ down to T$_{CDW}$= 110 K. The phonon gets overdamped upon cooling and the anomalies are broad in momentum space, identifying the electron-phonon interaction as the driving force of the electronically modulated structural instability. Density functional theory (DFT) calculations including non-perturbative anharmonic effects through the stochastic self-consistent harmonic approximation (SSCHA) \cite{PhysRevB.89.064302,PhysRevB.96.014111,PhysRevB.98.024106} can reproduce the temperature dependence of the soft mode and the T$_{CDW}$ onset only when the out-of-plane van der Waals interactions are considered. The results of our calculations show strongly momentum-dependent electron-phonon matrix-elements, matching the linewidth extracted experimentally, and corroborating the critical role of the EPI in the formation of the CDW.

High-quality single crystals of VSe$_2$ with dimensions 2$\times$2$\times$0.05 mm$^3$ were grown by chemical vapor transport (CVT) using iodine as transport agent. The high-resolution IXS experiments were carried out using the HERIX spectrometer at the 30-ID beamline of the Advanced Photon Source (APS), Argonne National Laboratory. The incident beam energy was 23.72 keV and the energy and momentum resolution was 1.5 meV and  0.7 nm$^{-1}$, respectively, \cite{Said20}. The components ($h$ $k$ $l$) of the scattering vector are expressed in reciprocal lattice units (r.l.u.), ($h$ $k$ $l$)= $h \mathbf{a}^*+k \mathbf{b}^*+l \mathbf{c}^*$, where $\mathbf{a}^*$, $\mathbf{b}^*$, and $\mathbf{c}^*$ are the reciprocal lattice vectors. The experimental lattice constants of the hexagonal unit cell at room temperature are \textit{a}= 3.346 \r{A}, \textit{c}= 6.096 \r{A}, and $\gamma$= 120$^{\circ}$. Here, we focus on the low energy acoustic phonon branches dispersing along the (0$<$\textit{h}$<$0.5 0 -0.3) direction in the Brillouin zone near the reciprocal lattice vector \textbf{G}$_{201}$, thus, in the range (2+\textit{h} 0 -0.3) with 0$<$\textit{h}$<$0.5. The variational SSCHA \cite{PhysRevB.89.064302,PhysRevB.96.014111,PhysRevB.98.024106} method was used to calculate temperature-dependent phonons fully accounting for non-perturbative anharmonic effects. The variational free energy minimization of the SSCHA was performed by calculating forces on 4$\times$4$\times$3 supercells (commensurate with \textit{q}$_{CDW}$) making use of DFT within the Perdew-Burke-Ernzerhof (PBE) \cite{PhysRevLett.77.3865} parametrization of the exchange-correlation functional. Van der Waals corrections were included within Grimme's semiempirical approach \cite{doi:10.1002/jcc.20495}. Harmonic phonon frequencies and electron-phonon matrix elements were calculated within density functional perturbation theory (DFPT) \cite{RevModPhys.73.515}. The force calculations in supercells needed for the SSCHA as well as the DFPT calculations were performed within the {\sc Quantum ESPRESSO} package~\cite{0953-8984-21-39-395502,0953-8984-29-46-465901} (see Supplementary Information for further details on the calculations
 \bibnote{See Supplementary Information for further details on the experimental procedure and calculations that includes citations to Refs. \cite{Marzari1,Marzari2,Wannier90}.}).  

\begin{figure}[h!]
\begin{center}
\includegraphics[width=\columnwidth,draft=false]{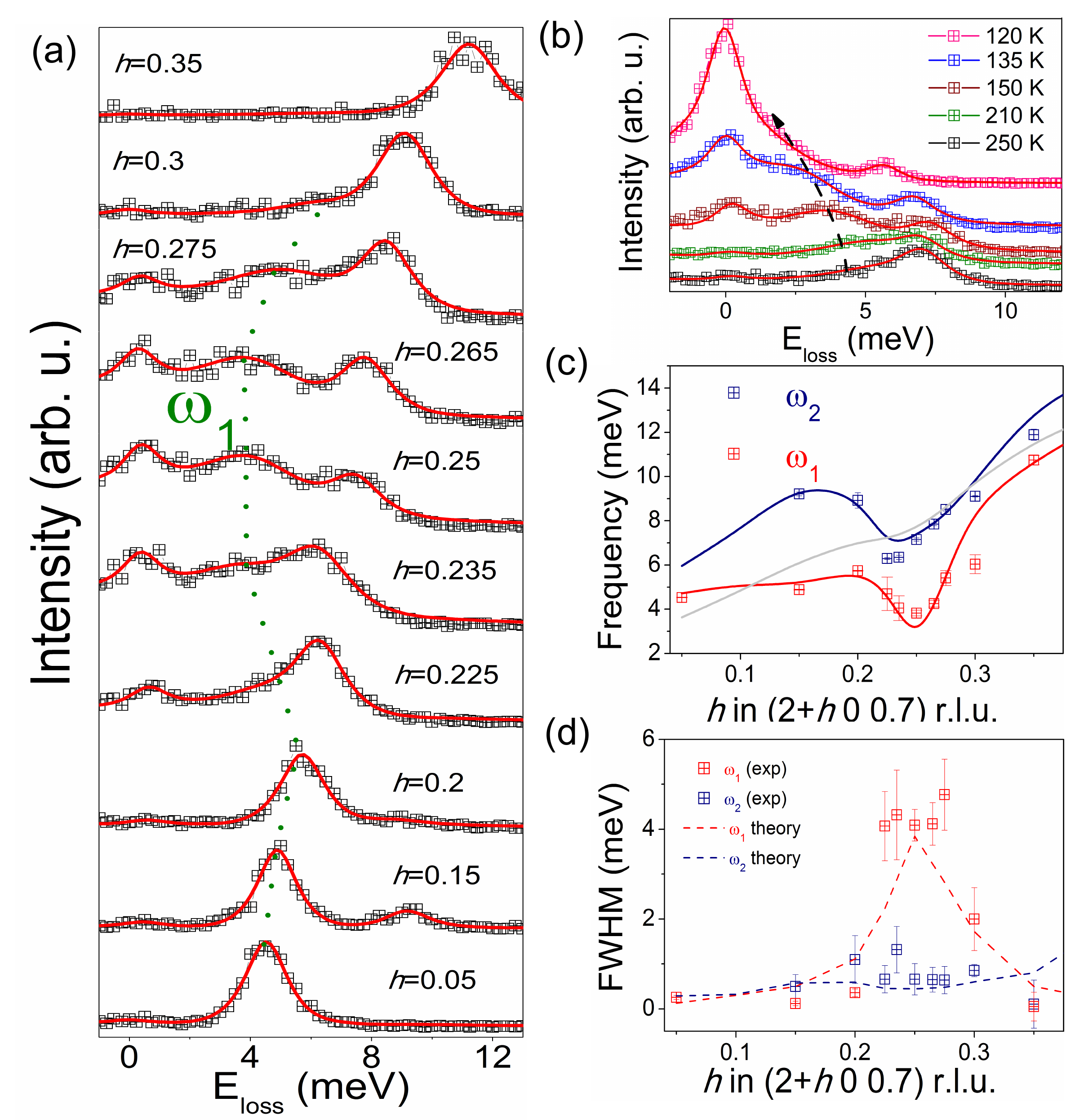}
\caption{(a) IXS energy-loss scans at (2+\textit{h} 0 0.7) r.l.u for 0.15$<$\textit{h}$<$0.35 and 150 K. The dotted green line follows the dispersion of the soft phonon mode, $\omega_1$ (see text). Each spectrum is vertically shifted for clarity. (b) Energy loss scans as a function of temperature at (2.25 0 0.7) r.l.u. The black arrow follows the softening of the low energy acoustic mode upon cooling. In both (a) and (b) red lines are the result of the fitting to damped harmonic oscillator functions convoluted with the instrumental resolution.  (c) Momentum dependence of the frequency  of the $\omega_1$ and $\omega_2$ branches at 150 K. The anharmonic phonon dispersions of the acoustic modes obtained at 150 K are plotted as solid lines. The grey line represents the acoustic mode that is silent in IXS. (d) Experimental (symbols) and theoretical (dashed lines) momentum dependence of the linewidth for $\omega_1$ and $\omega_2$. The theoretical calculation accounts for both the electron-phonon and anharmonic contributions to the linewidth.}
\label{Fig2}
\end{center}
\end{figure}

Fig.1(b) displays the temperature dependence of the elastic signal at the critical wavevector \textit{q}$_{CDW}$=(2.25 0 0.7) upon cooling from 300 K. The elastic line shows a sharp onset of the CDW superlattice peak at 110 K and a weak diffuse elastic \textit{central} peak due to low-energy critical fluctuations is visible below $\sim$ 140 K, implying little structural disorder. No indications of charge instabilities were observed along the $\Gamma\rightarrow$M and $\Gamma\rightarrow$L directions. The mean field critical exponent obtained in the disordered phase at T$>$T$_{CDW}$, $\gamma$ = 1.303 $\pm$ 0.004, is consistent with the existence of a 3D regime of critical fluctuations of an order parameter of dimensions \textit{n}=2, as expected for a classical \textit{XY} universality class \cite{Bak78}. A similar critical exponent has been observed in the quasi-1D conductor blue bronze K$_{0.3}$MoO$_3$ \cite{Gir89} and ZrTe$_3$ \cite{Hoe09}, which develops a giant Kohn anomaly at the CDW transition.

Fig. 1(c) displays the momentum dependence of the inelastic spectra at (2+\textit{h} 0 0.7) r.l.u. for 0.15$<$\textit{h}$<$0.45 at 300 K. Optical phonons appear above 17 meV and do not overlap with the acoustic branches. At all momentum transfers, 0$<$\textit{h}$<$0.5, the spectrum consists of 2 phonons, labeled $\omega_1$ and $\omega_2$ in Fig. 1(d), in good agreement with the results of the theoretical calculations (see supplementary information for a precise description and assignment of the 2 branches). The third acoustic mode is silent in IXS as its polarization vector is perpendicular to the wavevector. Both $\omega_1$ and $\omega_2$ belong to the same irreducible representation and, thus, do not cross. For \textit{h}$<0.2$, $\omega_1$ develops more spectral weight than $\omega_2$ and, for \textit{h}$>$0.2, the intensity of $\omega_2$ increases and $\omega_1$ leads an apparent asymmetric broadening of $\omega_2$, as depicted in the fig. 1d. To obtain quantitative information of the frequency and the phonon lifetime, the experimental scans were fitted using standard damped harmonic oscillator functions convoluted with the experimental resolution of $\sim$ 1.5 meV (see Fig. 1(d) and Supplementary Information for a detailed analysis of the fitting). The frequencies of the low energy acoustic branches $\omega_1$ and $\omega_2$ start around 4 and 8 meV, respectively, and end at $\sim$ 13 meV. Remarkably, the results of our \textit{ab initio} anharmonic SSCHA phonon calculations including van der Waals forces show that both $\omega_1$ and $\omega_2$ do not follow sinusoidal dispersion, but develop a dip at \textit{h}$\approx$ 0.25 r.l.u. The theoretical dispersion nicely matches the experimental data from the zone center to the border of the Brillouin zone (BZ), as shown in Fig. 1(e). In fact, the results of the harmonic phonon calculations indicate that the high temperature structure of 1\textit{T}-VSe$_2$ is unstable towards a CDW transition (see Fig. 4). It is clear, thus, that anharmonicity stabilizes 1\textit{T}-VSe$_2$ at high temperatures. On the other hand, the linewidth extracted from the analysis (Fig. 1(f), symbols) of the $\omega_2$ mode is resolution limited across the whole BZ. Nevertheless, the linewidth of the $\omega_1$ branch is no longer resolution limited between 0.2$<$\textit{h}$<$0.3 r.l.u. and develops an anomalously large broadening of $\sim$ 4 meV at \textit{h}=0.25 r.l.u. Again, the experimental broadening is well captured by our calculations (dashed lines in Fig. 1(f)), indicating that the large enhancement of the broadening is mainly due to the EPI even if the anharmonic contribution to the linewidth also peaks at \textit{h}=0.25 r.l.u. (supplementary information).

Given the observation of the phonon broadening at room temperature and the good agreement between theory and experiment, we proceed with the analysis of the lattice dynamics at lower temperatures. At 250 K, the phonon with energy $\sim$ 7 meV ($\omega_2$) shows a clear asymmetric broadening at \textit{q}$_{CDW}$, i.e, the corresponding branch $\omega_1$ appears to develop a redshift as a function of temperature (fig. 2(b)). The dispersion of $\omega_2$ at 150 K is similar to the one at 300 K. Contrarily, $\omega_1$ lowers its energy, softening from room temperature down to 110 K. The softening extends over a wide region of momentum space 0.225$<$\textit{h}$<$0.3 r.l.u. (0.15 \r{A}$^{-1}$) at 150 K, see green dotted line in fig. 2(a). The pronounced instability of this acoustic mode and its broad extension in momentum space are consistent with the results of our anharmonic phonon calculations (solid lines in fig. 2(c)). The momentum space spread of the softening indicates a substantial localization of the phonon fluctuations in real space due to the EPI, questioning the pure nesting mechanism suggested by ARPES \cite{Stro12}. More importantly, the softening of this branch represents the first indication of the lattice response to the formation of the 3D-CDW in VSe$_2$. The analysis of the linewidth reveals that the lifetime of $\omega_2$ remains nearly constant across the BZ and is resolution limited (fig. 2(d)). On the other hand, the softening of the $\omega_1$ mode at 150 K is accompanied by an enhancement of the linewidth, as shown in fig. 2(d) (6 meV linewidth at 120 K, fig. 3(f)) and, again, well modelled by the \textit{ab initio} calculations (dashed lines in fig. 2(d)). 

\begin{figure}[h!]
\begin{center}
\includegraphics[width=1.0\columnwidth,draft=false]{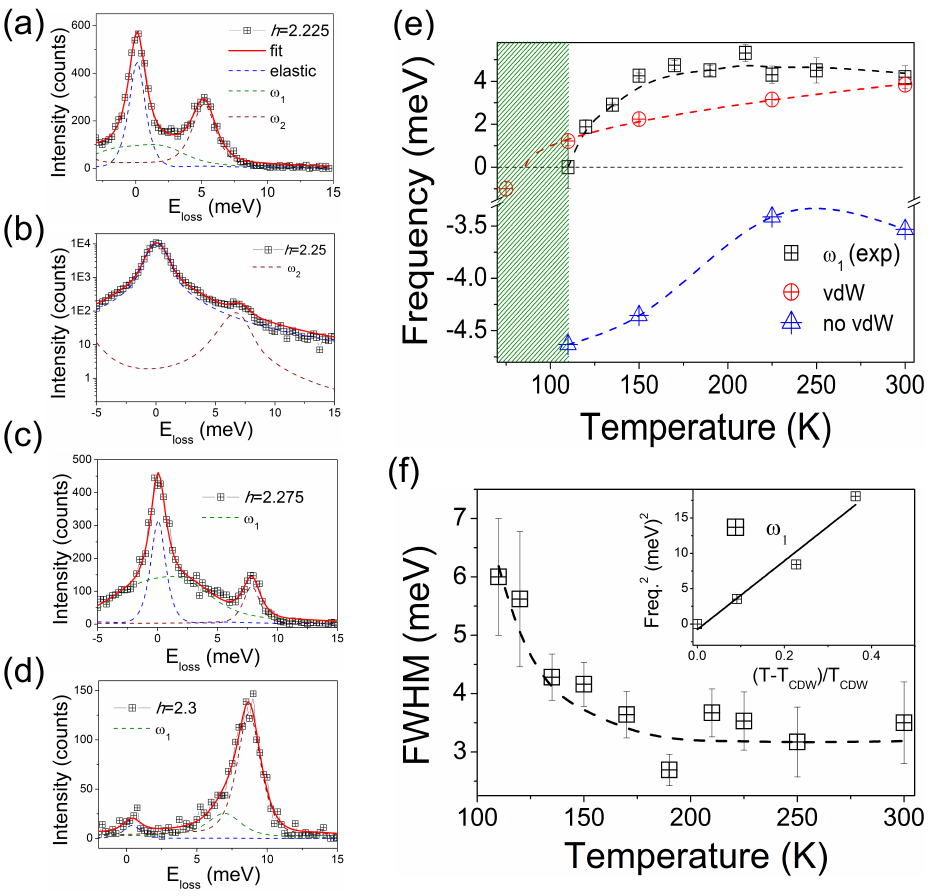}
\caption{(a-d) Representative IXS spectra at 110 K and their corresponding fitting. The IXS scan at \textit{h}=2.25 r.l.u. is presented in logarithmic scale due to the large enhancement of the elastic line. $\omega_1$ stands for the soft mode. (e) Temperature dependence of the energy of the $\omega_1$ branch and the anharmonic theoretical frequencies obtained with and without van der Waals corrections. The shaded area defines the CDW region. (f) Temperature dependence of the linewidth. Inset, squared frequency of the soft mode as a function of the reduced temperature. Lines are guides to eye.}
\label{fran_p}
\end{center}
\end{figure}

At the critical temperature, T$_{CDW}$=110 K, the spectrum is dominated by an elastic central peak at zero energy loss (FWHM= 0.05 r.l.u. and $\Delta$E= 1.6 meV), thus, the soft mode is no longer resolvable (see fig. 3(a-d)). Fig. 3(e) displays the temperature dependence of the soft mode, $\omega_1$, as well as the frequency of the phonon obtained \textit{ab initio} with and without including van der Waals corrections. Our anharmonic calculations, which predict that the $\omega_1$ frequency vanishes between 75 and 110 K, are in rather good agreement with the experimentally measured phonon frequencies and the CDW temperature onset, T$_{CDW}$=110 K. When the SSCHA anharmonic calculation is repeated without including the van der Waals corrections (blue triangles in fig. 3(e)), the softest acoustic mode at \textit{q}$_{CDW}$ remains unstable even at room temperature. Remarkably, the weak van der Waals forces (of the order of $\sim$1mRy/$a_0$ for a typical SSCHA supercell calculation) are responsible for the stabilization of the 1\textit{T} structure of VSe$_2$ and play a crucial role in melting the CDW. On the other hand, the damping ratio, $\Gamma/\tilde{\omega}_q$, increases upon cooling and the phonon becomes critically overdamped at \textit{q}$_{CDW}$ and 110 K \bibnote{The damping ratio $\Gamma/\tilde{\omega}_q$ is given by $\omega_0=(\tilde\omega_q^2-\Gamma^2)^{1/2}$, where $\Gamma$ is the linewidth $\tilde\omega_q$ is the phonon energy renormalized by the real part of the susceptibility and $\omega_0$ is the energy of the phonon fitted to damped harmonic oscillator}. The critical exponent derived from the fitting of the phonon frequency \textit{vs} reduced temperature ((T-T$_{CDW}$)/T$_{CDW}$), $\beta$=0.52$\pm$0.4, agrees with the square-root power law expected from the mean field theory (inset of fig. 3(f)). 
%Similar critical exponent was observed in 1D Peierls system ZrTe$_3$ \cite{Hoe09} and K$_2$Pt(CN)$_4$Br$_{0.3}$\dot3H$_2$0 \cite{Com73}, where the mean-field transition is located at much higher temperature than the Peierls transition, T$_{MF}$\approx4\times T$_P$ \cite{Lee73}. 

\begin{figure}[h!]
\begin{center}
\includegraphics[width=\columnwidth,draft=false]{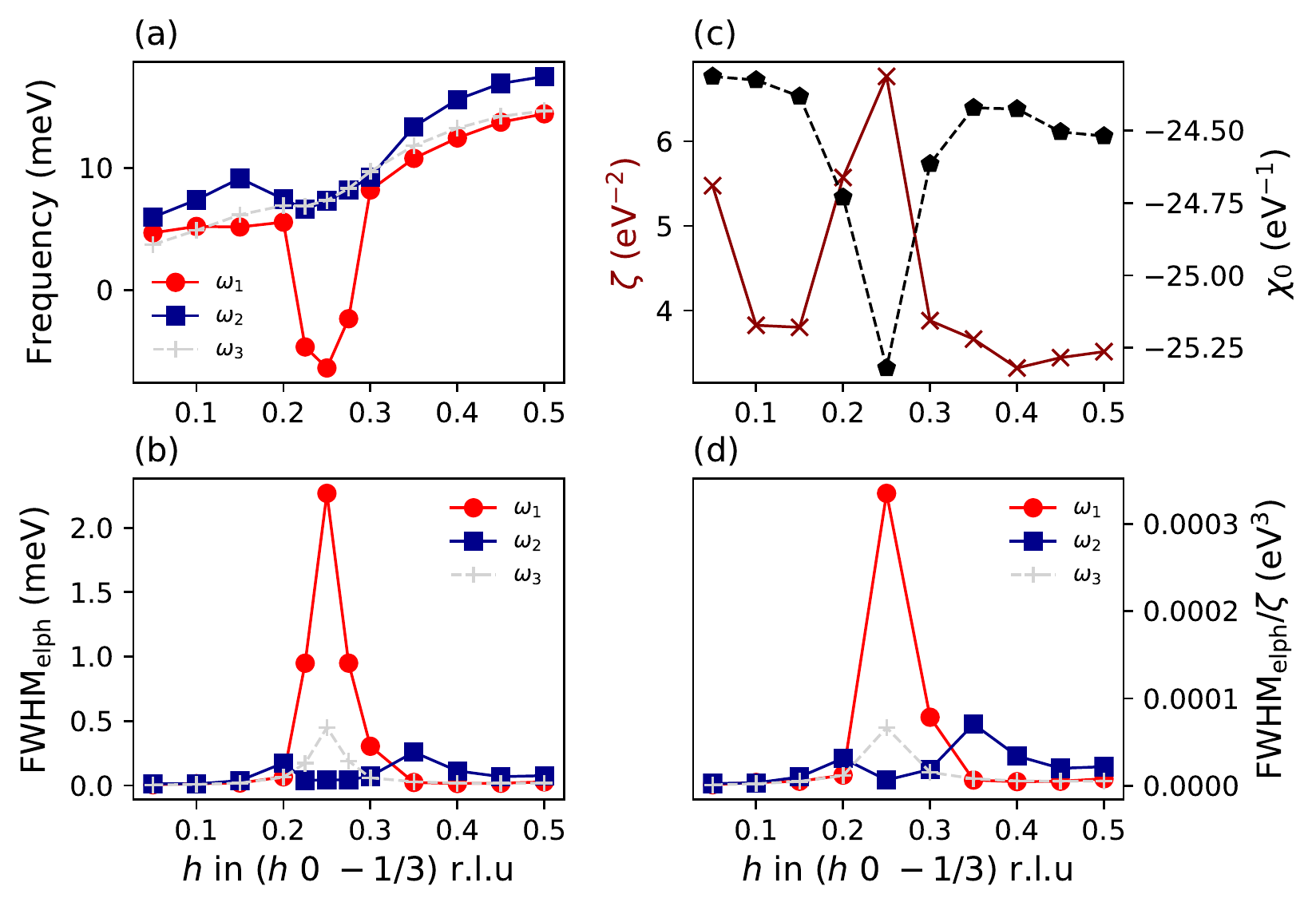}
\caption{(a) Calculated harmonic phonon spectra of 1\textit{T}-VSe$_2$ along (\textit{h} 0 -1/3). Only acoustic modes are shown. The grey line denotes the mode silent in IXS, which is labeled as $\omega_3$ here. (b) Phonon linewidth (full width at half maximum) given by the electron-phonon interaction for the same modes. (c) Real part of the non-interacting susceptibility, $\chi_0$, as well as the nesting function, $\zeta$, at the same wavevectors. (d) Ratio between the full width at half maximum given by the electron-phonon interaction and the nesting function.
}
\label{Fig4}
\end{center}
\end{figure}

%$\chi_0 = 1/N_k\sum_{\mathbf{k}nm}1/(\varepsilon_{n\mathbf{k}}-\varepsilon_{m\mathbf{k}+\mathbf{q})$

Having achieved a comprehensive description of the CDW and its temperature dependence, we address the crucial role of the EPI and nesting mechanism in the formation of the charge modulated state. In Fig. \ref{Fig4}, we plot the calculated harmonic phonon frequency together with the electron-phonon linewidth of the three acoustic modes along \textit{q}=(\textit{h} 0 -1/3) calculated within DFPT. As it can be seen, the harmonic phonon instability of $\omega_1$ coincides with a huge increase of its linewidth associated with the EPI. The softening and the increase of the electron-phonon linewidth specially affect the $\omega_1$ mode, which suggests that the electron-phonon matrix elements are strongly mode and momentum dependent and have an strong impact on the real part of the phonon self-energy, which determines the harmonic phonon frequencies \cite{PhysRevB.80.241108,PhysRevB.77.165135}. This behavior is similar to the one reported for 1\textit{T}-TiSe$_2$ and 2\textit{H}-NbSe$_2$ \cite{Web11Ti,Web11Nb}. The real part of the non-interacting susceptibility $\chi_0(\mathbf{q})$, which is calculated with constant matrix elements (see Supplementary Information) but captures the full Fermi surface topology and also affects the real part of the phonon self-energy, has a softening of around 4\%  at \textit{q}$_{CDW}$, which seems insufficient to explain the large softening of the $\omega_1$ mode. The nesting function $\zeta(\mathbf{q})$ (supplementary information) peaks at \textit{q}$_{CDW}$, which indicates that the CDW vector coincides with a nested region of the Fermi surface. However, the phonon linewidth of the $\omega_1$ mode coming from the EPI depends more drastically on momentum than the nesting function: it changes by orders of magnitude as a function of $\mathbf{q}$ while the nesting function only by less than a factor of two. This is highlighted in the ratio between the linewidth and the nesting function plotted on fig. \ref{Fig4}(d), which measures the momentum dependence of the electron-phonon matrix elements. This ratio depends much more strongly on momentum than the nesting function itself and resembles the linewidth dependence. This result supports that the EPI is the main driving force of the CDW transition in 1\textit{T}-VSe$_2$ despite the presence of nesting. Nevertheless, the \textbf{q}-range over which the phonon softens, $\Delta$\textbf{q}$\approx$ 0.075 r.l.u., is a factor of 3 less than in 1\textit{T}-TiSe$_2$ \cite{Web11Ti}, where EPI and excitonic correlations are responsible for the structural instability and the CDW order, pointing to an intricate relationship between EPI and Fermi surface nesting scenarios in VSe$_2$. %Indeed, the nearly flat electronic susceptibility, \chi$_0$, we have calculated is independent of \textit{h} and \textit{l}, (figure 4(c)), and may cause imperfect nesting along \textit{l}, where part of the Fermi surface remains ungapped, leading to a certain strength of electron-phonon coupling constant and phonon renormalization \cite{Ichi90}. 

%We would like to end the discussion by pointing out the similarities of the weak out-of-plane interactions in driving the phonon renormalizations in superconducting cuprates. It has been observed that the 3D-CDW in 123 YBa$_2$Cu$_3$O$_{6+\delta}$ is stabilized through the hybridization of the Cu chains with Cu-O planes \cite{Blu18} and a very strong phonon softening is observed \cite{Kim18}. Indeed, low energy phonon broadening has also been reported in the van der Waals cuprate Bi$_2$Sr$_2$CaCu$_2$O$_{8+\delta}$ \cite{He18}, demonstrating the crucial role out-of-plane coupling of cuprate layers.  

In conclusion, we have observed with high resolution IXS that the CDW  transition in 1\textit{T}-VSe$_2$ is driven by the collapse of an acoustic mode at \textit{q}$_{CDW}$=(0.25 0 -0.3) exactly at T$_{CDW}$=110 K. The high-temperature 1\textit{T}-VSe$_2$ phase is stable thanks to anharmonic effects. The observed  wide softening in momentum space, the calculated strongly momentum dependent electron-phonon linewidth that peaks at \textit{q}$_{CDW}$, and the weaker dependence on the wavevector of the susceptibility suggest that the EPI is the main driving force of the CDW transition despite the presence of nesting. Moreover, the results show that van der Waals forces are responsible for the melting of the CDW. We attribute the dominant role of van der Waals forces here to the out-of-plane nature of the CDW, which is in contrast to, for instance, 2\textit{H}-NbSe$_2$, where the CDW implies an in-plane modulation and the bulk and monolayer transition temperatures seem to be similar \cite{Ugeda2016,bianco2020weak}. Indeed, our anharmonic phonon calculations straightforwardly explain the enhancement of the CDW in monolayer VSe$_2$, T$_{CDW}$= 220 K \cite{PhysRevLett.121.196402}, where the out-of-plane van der Waals interactions are absent. The critical role of out-of-plane coupling of layers has also been highlighted in cuprate superconductors \cite{Blu18,Kim18,He18}.

We acknowledge valuable discussions with V. Pardo, A. O. Fumega and M. Hoesch. S.B-C thanks the MINECO of Spain through the project PGC2018-101334-A-C22. F.M. and L.M. acknowledge support by the MIUR PRIN-2017 program, project number 2017Z8TS5B. M.C. acknowledges support from Agence Nationale de la Recherche, Project ACCEPT, Grant N. ANR-19-CE24-0028 and M.C and F.M. the Graphene Flagship Core 3. Calculations were performed at the Joliot Curie-AMD supercomputer under the PRACE project RA4956.

%\bibliography{VSe2}


\end{document}